# The Butterfly Effect of Technology: How Various Factors accelerate or hinder the Arrival of Technological Singularity

Hooman Shababi
Faculty Member at Rahedanesh Institute of Higher Education, Babol, Mazandaran, Iran
Email: Hooman-shababi@rahedanesh.ac.ir
Cellphone: +989116419188
Fax: +981132379892

Abstract
This article explores the concept of technological singularity and the factors that could accelerate or hinder its arrival. The butterfly effect is used as a framework to understand how seemingly small changes in complex systems can have significant and unpredictable outcomes. In section II, we discuss the various factors that could hasten the arrival of technological singularity, such as advances in artificial intelligence and machine learning, breakthroughs in quantum computing, progress in brain-computer interfaces and human augmentation, and development of nanotechnology and 3D printing. In section III, we examine the factors that could delay or impede the arrival of technological singularity, including technical limitations and setbacks in AI and machine learning, ethical and societal concerns around AI and its impact on jobs and privacy, lack of sufficient investment in research and development, and regulatory barriers and political instability. Section IV explores the interplay of these factors and how they can impact the butterfly effect. Finally, in the conclusion, we summarize the key points discussed and emphasize the importance of considering the butterfly effect in predicting the future of technology. We call for continued research and investment in technology to shape its future and mitigate potential risks.

## 1- Introduction

Technological singularity is a hypothetical future event in which artificial intelligence surpasses human intelligence and becomes capable of recursive self-improvement, leading to an exponential increase in technological progress. This concept was first introduced by mathematician and computer scientist Vernor Vinge in his 1993 essay "The Coming Technological Singularity: How to Survive in the Post-Human Era" (Vinge, 1993). Since then, it has been the subject of much debate and discussion within the scientific and technological communities.

The idea of technological singularity is based on the notion that as artificial intelligence becomes more advanced, it will eventually become capable of improving itself, leading to a rapid increase in its capabilities (Kurzweil, 2005). This self-improvement could lead to an intelligence explosion, where AI becomes so advanced that it surpasses human intelligence and becomes capable of solving problems and creating innovations that humans are unable to comprehend.

One of the key features of technological singularity is the idea of accelerating returns. This means that as technology becomes more advanced, its rate of progress increases, leading to an exponential increase in its capabilities (Kurzweil, 2001). This could potentially lead to a runaway effect, where technology becomes so advanced that it becomes difficult or impossible to predict its impact on society.

There are several different approaches to defining technological singularity. Some researchers focus on the technological aspects, such as advances in machine learning, artificial intelligence, and robotics (Bostrom, 2014). Others take a more philosophical approach, examining the implications of the singularity for human society, ethics, and morality (Hawking, Tegmark, & Wilczek, 2014).

One of the key challenges in defining technological singularity is that it is a speculative concept that is based on assumptions about the future development of technology. There is no way to know for certain whether the singularity will occur or what its impact will be. However, many researchers believe that it is a possibility worth exploring, given the potential risks and benefits that it presents (Bostrom, 2014).

One way to explore the concept of technological singularity is through the use of thought experiments. For example, the philosopher Nick Bostrom has proposed the simulation argument, which suggests that if it is possible to create highly realistic computer simulations of conscious beings, it is likely that we are already living in such a simulation. This argument has implications for the possibility of post-human civilizations and the potential for technological singularity (Bostrom, 2003).

The concept of technological singularity is a crucial topic in science, technology, and philosophy. It is essential to understand the factors that affect the arrival of technological singularity, as it has the potential to dramatically alter the course of human history. The development of super intelligent AI could lead to significant advances in fields such as medicine, energy, and space exploration. However, it could also have negative consequences if we are not prepared for the potential risks associated with advanced AI, such as job displacement, inequality, and existential risks (Bostrom, 2014).

By understanding the factors that affect the arrival of technological singularity, we can take steps to mitigate these risks and ensure that the benefits of advanced AI are shared equitably across society. For example, we can invest in education and training programs to prepare workers for the jobs of the future, implement policies to promote diversity and inclusion in AI development, and establish ethical guidelines for the use of AI (Floridi, Cowls, Beltrametti, Chatila, Chazerand, Dignum, Luetge, Madelin, Pagallo, Rossi, Schafer, & Valcke, 2018).

Moreover, understanding the factors that affect the arrival of technological singularity can help us make more informed decisions about how to shape the future of technology. For instance, we can make strategic investments in areas such as AI research, quantum computing, and biotechnology to accelerate the development of advanced technologies. We can also foster collaboration and knowledge sharing among scientists, engineers, and other stakeholders to promote innovation and progress (Kurzweil, 2005).

Additionally, understanding the factors that affect the arrival of technological singularity can help us better understand the nature of intelligence and consciousness. The development of super intelligent AI could have profound implications for our understanding of these concepts and how they relate to human experience. By exploring these topics, we can deepen our understanding of what it means to be human and how we can harness the power of technology to enhance our lives (Bostrom, 2014).

The butterfly effect is a term coined to describe the idea that small changes in one system can have significant effects on a complex system. It originates from chaos theory, where it refers to the sensitivity of initial conditions in a non-linear system. The term became popular in the 1960s when meteorologist Edward Lorenz used it to describe how a small change in the initial conditions of a weather system, such as the flap of a butterfly's wings, could lead to a significant difference in the eventual outcome of the weather (Lorenz, 1963).

In the context of technology, the butterfly effect refers to the idea that small changes in the development of a technology can have significant and unforeseeable consequences. For example, the invention of the internet, which started as a way for researchers to share information, has had a profound impact on society and the economy, enabling new industries, business models, and forms of communication (Crawford, 2016).

The butterfly effect is relevant to technology in several ways. First, it highlights the importance of understanding the potential consequences of new technologies before they are widely adopted. Small changes in the design, development, or deployment of a technology can have far-reaching effects that are difficult to predict. For example, the unintended consequences of social media platforms, such as the spread of misinformation and polarization, have highlighted the need for greater attention to the potential social impacts of new technologies (Christakis & Fowler, 2009).

Second, the butterfly effect underscores the importance of collaboration and knowledge sharing in technology development. Because of the complexity of technological systems, it is often challenging to anticipate all the potential outcomes of a particular technology. Collaboration among experts from different disciplines can help identify potential risks and unintended consequences of a new technology and develop solutions to mitigate them (Mackenzie & Wajcman, 1999).

Third, the butterfly effect highlights the importance of ethical considerations in technology development. Because of the potential for unforeseen consequences, it is important to consider the potential impact of a new technology on society, the environment, and individual rights and freedoms. Ethical considerations should be integrated into the design, development, and deployment of new technologies to ensure that they align with societal values and goals (Floridi et al., 2018).

In conclusion, understanding the factors that affect the arrival of technological singularity and the potential impact of technological developments is crucial for ensuring a responsible and sustainable future. The butterfly effect, as a concept, highlights the importance of taking into account the potential consequences of small changes and the need for collaboration and ethical considerations in technology development. In the following sections, this article will explore the various factors that influence the arrival of technological singularity, including technological progress, social and economic factors, ethical considerations, and regulatory frameworks. By examining these factors, we can gain a better understanding of the potential impact of technological singularity and develop strategies to manage its arrival.

**2- Factors that could accelerate the arrival of technological singularity**

**2-1-Advances in artificial intelligence and machine learning**

Advances in artificial intelligence and machine learning are among the key drivers of technological singularity. Artificial intelligence (AI) refers to the development of computer systems that can perform tasks that usually require human intelligence, such as perception, reasoning, learning, and decision-making (Russell & Norvig, 2020). Machine learning, on the other hand, is a subset of AI that involves training algorithms on large datasets to learn patterns and make predictions or decisions without being explicitly programmed (Alpaydin, 2010).

AI and machine learning have already transformed various industries, such as healthcare, finance, and transportation, and are expected to continue to do so in the future (Marr, 2018). As AI systems become more advanced, they may even surpass human intelligence, leading to the arrival of technological singularity.

One significant breakthrough in AI is deep learning, which involves training neural networks on vast amounts of data to perform complex tasks, such as image and speech recognition, natural language processing, and decision-making (LeCun, Bengio, & Hinton, 2015). Deep learning has led to remarkable advancements in fields such as computer vision, robotics, and autonomous vehicles (Goodfellow, Bengio, & Courville, 2016). For instance, self-driving cars use deep learning algorithms to recognize and respond to their environment, making them safer and more efficient than traditional vehicles (Bojarski et al., 2016).

Another area of AI that has the potential to impact the arrival of technological singularity is general artificial intelligence (AGI). AGI refers to the development of machines that can perform any intellectual task that a human can (Bostrom, 2014). While current AI systems are designed to perform specific tasks, such as playing chess or diagnosing diseases, AGI would have the ability to reason, learn, and solve problems in a more general sense. The development of AGI would represent a significant milestone towards technological singularity.

However, as AI and machine learning continue to advance, there are concerns about the potential risks and ethical implications of their development (Bostrom, 2014). For example, AI systems could be used for malicious purposes, such as cyber attacks or autonomous weapons (Bryson, 2018). Additionally, there are concerns about job displacement, as machines may be able to perform tasks more efficiently and cost-effectively than humans (Frey & Osborne, 2017). It is important to consider these ethical and social implications as we continue to develop AI and machine learning systems.

In summary, advances in AI and machine learning are among the key factors driving the arrival of technological singularity. These developments have already transformed various industries and have the potential to lead to significant advancements in the future. However, it is crucial to consider the potential risks and ethical implications of AI and machine learning development to ensure a responsible and sustainable future.

## 2-2-Breakthroughs in quantum computing

Quantum computing is an emerging technology that is set to revolutionize various fields, including finance, logistics, and scientific research. Unlike classical computing that uses bits, which can represent only 0s or 1s, quantum computing uses qubits, which can represent both 0s and 1s simultaneously due to their superpositioning. The use of qubits in quantum computing enables machines to perform complex calculations at a much faster rate than classical computers.

The development of quantum computing has been an ongoing process for many years, and recent breakthroughs have brought us closer to achieving quantum supremacy, a term coined by John Preskill in 2012 to describe the point at which a quantum computer can perform a calculation that no classical computer can match (Preskill, 2018).

One of the recent breakthroughs in quantum computing was achieved by Google's research team in 2019. They demonstrated quantum supremacy by using a quantum processor called Sycamore to solve a problem in just 200 seconds that would take a classical computer 10,000 years to solve (Arute et al., 2019). This achievement has opened up new possibilities in scientific research and cryptography.

Another significant breakthrough in quantum computing is the development of error correction codes, which can mitigate the errors that arise due to the fragile nature of qubits. Microsoft's research team announced in 2020 that they had developed a topological qubit that could enable the creation of fault-tolerant quantum computers, which can perform calculations with minimal errors (Huang, 2020).

These breakthroughs have made quantum computing more practical, and several companies, including IBM, Google, and Microsoft, are investing heavily in this technology. Quantum computing has the potential to solve some of the most challenging problems in science and industry, such as the simulation of complex chemical reactions and optimization problems in logistics and finance.

Another important breakthrough in quantum computing is the development of quantum algorithms, which are designed to solve specific problems much faster than classical algorithms. One example of a quantum algorithm is Shor's algorithm, which can efficiently factor large numbers, a problem that is intractable for classical computers. This algorithm has significant implications for cryptography because it can break many of the cryptographic protocols that are currently in use (Shor, 1994).

In addition to its potential for solving problems that are currently infeasible for classical computers, quantum computing also has the potential to revolutionize the field of artificial intelligence. Quantum machine learning, which combines the power of quantum computing with machine learning algorithms, can improve the speed and accuracy of tasks such as pattern recognition and data analysis (Biamonte et al., 2017).

Despite the promising developments in quantum computing, there are still significant challenges that need to be overcome before it can become a mainstream technology. One of the main challenges is the need for a large number of qubits, as quantum computers require thousands of qubits to perform useful calculations. Another challenge is the need for error correction, as the fragile nature of qubits makes them susceptible to errors.

Despite these challenges, quantum computing is a rapidly growing field, and its potential for solving complex problems has attracted significant interest from researchers and companies alike. The development of quantum computing could potentially accelerate the arrival of technological singularity, as it could enable machines to perform calculations and make decisions at a much faster rate than humans.

**2-3- Progress in brain-computer interfaces and human augmentation**

Brain-computer interfaces (BCIs) are devices that allow direct communication between the brain and an external device, such as a computer or a prosthetic limb. In recent years, there has been significant progress in the development of BCIs, which could have significant implications for human augmentation and the arrival of technological singularity.

One area where BCIs could have significant impact is in the treatment of neurological disorders. For example, BCIs could be used to restore motor function to individuals with spinal cord injuries or to alleviate symptoms of Parkinson's disease (Gilja et al., 2015). BCIs could also be used to improve communication for individuals with locked-in syndrome, a condition in which an individual is fully conscious but unable to move or communicate (Vansteensel et al., 2016).

BCIs also have the potential to enhance human cognitive abilities, such as memory and learning. One approach to cognitive enhancement is through the use of neural prosthetics, which are devices that can be implanted in the brain to enhance cognitive function. For example, researchers have successfully implanted neural prosthetics in rats to enhance their memory (Berger et al., 2011). This technology could potentially be used to enhance human memory or to treat cognitive disorders such as Alzheimer's disease.

Another area where BCIs could have significant impact is in the development of human-machine interfaces. BCIs could be used to control devices such as prosthetic limbs or even to control autonomous vehicles or drones (Wang et al., 2020). BCIs could also be used to enhance the performance of individuals in high-stress environments, such as military or emergency responders.

The development of BCIs raises important ethical and social considerations, particularly around issues of privacy and autonomy. As BCIs become more sophisticated, they may allow individuals to communicate and interact with technology in ways that are not currently possible. This could raise concerns around the potential loss of privacy or control over one's own thoughts and actions (Clausen, 2016).

Despite these concerns, the development of BCIs has significant potential for human augmentation and the arrival of technological singularity. BCIs could enable humans to interact with technology in new and unprecedented ways, potentially leading to a merging of human and machine intelligence. The development of BCIs is a rapidly evolving field, and it will be important to continue to monitor and address the ethical and social implications of this technology.

## 2-4- Development of nanotechnology and 3D printing

Nanotechnology and 3D printing are two technological advancements that have the potential to revolutionize the way we manufacture and produce goods. Nanotechnology is the study of materials on a molecular or atomic scale, while 3D printing is the process of creating three-dimensional objects from a digital file. Both of these technologies have the potential to disrupt traditional manufacturing processes and enable the creation of complex structures and devices that were previously impossible to produce.

Nanotechnology has numerous applications in various fields such as electronics, medicine, and energy (Zheng et al., 2020). In the field of electronics, nanotechnology has enabled the production of smaller, faster, and more efficient devices. For example, nanowires and nanotubes have been used to create faster and more energy-efficient transistors (Lee et al., 2018). In medicine, nanotechnology has led to the development of new drug delivery systems, such as nanoparticles that can target specific cells or tissues in the body (Wang et al., 2019). Nanotechnology has also enabled the development of new materials with unique properties, such as graphene, which is a strong, lightweight, and highly conductive material (Li et al., 2019).

3D printing, on the other hand, has already transformed the manufacturing industry by enabling the production of complex and customized parts with ease. 3D printing allows for the creation of parts with intricate geometries that would be difficult or impossible to produce with traditional manufacturing methods (Chen et al., 2019). It also enables the production of small batches of parts, reducing the need for costly tooling and setup. In addition, 3D printing has the potential to reduce waste and improve sustainability by enabling the production of parts with minimal material waste (Lindner et al., 2017).

However, there are also challenges that need to be addressed in order to fully realize the potential of nanotechnology and 3D printing. For example, the safety of nanomaterials needs to be carefully evaluated in order to avoid unintended consequences (Wu et al., 2019). In addition, the high cost of 3D printing equipment and materials can be a barrier to widespread adoption of the technology (Mitsouras et al., 2017).

Overall, the development of nanotechnology and 3D printing has the potential to significantly impact the manufacturing industry and various fields. As these technologies continue to advance and become more accessible, it will be important to carefully consider the benefits and challenges associated with their use.

### 3- Factors that could hinder the arrival of technological singularity

**3-1-Technical limitations and setbacks in AI and machine learning**
Advancements in AI and machine learning have brought numerous benefits, but there are also technical limitations and setbacks to these technologies. One of the biggest challenges is the "black box" problem. Machine learning models can produce accurate results, but it can be difficult to understand how they arrived at those results. This can lead to difficulties in troubleshooting and auditing, particularly in high-stakes applications such as medicine and finance (Kleinberg, Ludwig, Mullainathan, & Sunstein, 2018).

Another major limitation is the reliance on data quality. Machine learning models require large amounts of data to train and improve their performance. However, if the data is biased or incomplete, it can result in biased or inaccurate predictions. This is particularly problematic in fields such as criminal justice and hiring, where biased data can perpetuate discrimination (Angwin et al., 2016; Buolamwini & Gebru, 2018).

There are also technical limitations in the hardware used to support AI and machine learning. Traditional computer processing units (CPUs) are not well-suited to the massive parallel processing required for many AI applications. Graphics processing units (GPUs) have been developed specifically for this purpose and have greatly improved the speed and efficiency of machine learning models (Krizhevsky, Sutskever, & Hinton, 2017). However, even GPUs have limitations, particularly in terms of energy consumption and heat generation.

Another limitation is the high computational cost of deep learning, a type of machine learning that involves training neural networks with multiple layers of interconnected nodes. Deep learning models require vast amounts of processing power and memory, and training can take weeks or even months (Amodei et al., 2016). This limits the scalability of deep learning models and makes it difficult to apply them to real-time applications.

Despite these technical limitations and setbacks, researchers continue to develop new techniques and technologies to improve the performance and efficiency of AI and machine learning. For example, neuromorphic computing aims to mimic the architecture and function of the human brain, potentially overcoming some of the limitations of traditional computing hardware (Indiveri et al., 2011). Additionally, research into explainable AI seeks to develop models that can provide clear and interpretable explanations of their decision-making processes, improving transparency and accountability (Lipton, 2018).

### 3-2- Ethical and societal concerns around AI and its impact on jobs and privacy

As artificial intelligence (AI) becomes increasingly ubiquitous in our daily lives, there are growing concerns about its ethical and societal implications. In particular, there are concerns about the impact of AI on employment and privacy.

One of the most pressing concerns is the potential displacement of jobs by AI. As AI technology advances, machines are becoming capable of performing tasks that were once the exclusive domain of humans. This has led to fears that large numbers of jobs will be lost to automation, particularly in industries such as manufacturing and transportation (Frey & Osborne, 2017).

While some argue that AI will create new job opportunities, others warn that the transition to a more automated workforce could be difficult and disruptive, particularly for low-skilled workers who may lack the skills necessary to adapt to new roles (Autor, 2015). Additionally, there are concerns about the impact of AI on income inequality and the potential for a "winner-takes-all" economy in which a small number of highly skilled workers benefit at the expense of everyone else (Brynjolfsson & McAfee, 2014).

Privacy is another major concern surrounding AI. As AI becomes more advanced and ubiquitous, it has the potential to collect vast amounts of data about individuals, including their behaviors, preferences, and even their emotions. This has led to concerns about the potential for mass surveillance and the infringement of privacy rights (Scherer & Mueller, 2019).

Moreover, AI systems are only as unbiased as the data they are trained on. If the data is biased or incomplete, this can lead to algorithmic discrimination, which has already been documented in various domains, including criminal justice, hiring, and credit scoring (Barocas & Selbst, 2016; Buolamwini & Gebru, 2018; Eubanks, 2018). This can perpetuate social inequalities and undermine the legitimacy of AI systems.

Another ethical concern is the potential for AI to be used in ways that are harmful to society. For example, autonomous weapons powered by AI could be used to conduct warfare without human intervention, raising concerns about the risk of accidental or malicious use (Scharre, 2018). Additionally, the use of AI in surveillance and law enforcement raises concerns about the potential for abuse and the erosion of civil liberties (Holt & Calo, 2018).

To address these ethical and societal concerns, there have been calls for greater regulation and oversight of AI development and deployment. In particular, there is a need for greater transparency and accountability in the development and use of AI systems (Floridi et al., 2018). This includes ensuring that AI systems are transparent and explainable, so that their decision-making processes can be understood and audited.

There is also a need for ethical frameworks and guidelines to govern the development and deployment of AI. These frameworks should take into account the potential impact of AI on individuals and society, and ensure that AI is developed and used in a way that is consistent with human values and rights (Asaro, 2012). Furthermore, there is a need for greater education and public engagement around the ethical and societal implications of AI, so that individuals and communities can make informed decisions about the use of AI in their lives (Mittelstadt et al., 2016).

In conclusion, while AI has the potential to bring numerous benefits, there are also ethical and societal concerns surrounding its development and deployment. These concerns include the potential displacement of jobs, infringement of privacy, algorithmic discrimination, and the potential for AI to be used in harmful ways. To address these concerns, there is a need for greater regulation, transparency, and accountability in the development and use of AI, as well as the development of ethical frameworks and guidelines to govern the use of AI.

### 3-3- Lack of sufficient investment in research and development

The importance of research and development (R&D) in driving innovation and economic growth is widely recognized. However, there are concerns that many countries are not investing enough in R&D, particularly in emerging technologies such as artificial intelligence (AI). This lack of investment can have significant consequences, including a slowdown in the pace of technological progress, reduced competitiveness, and missed opportunities for economic growth.

One reason for the lack of investment in R&D is the perception that it is expensive and risky. Developing new technologies requires significant resources and involves a high degree of uncertainty. However, the benefits of R&D can be significant, both in terms of economic growth and in addressing societal challenges such as climate change and healthcare.

Governments have an important role to play in funding R&D, particularly in emerging technologies where private sector investment may be limited. In many countries, public funding for R&D has been declining in recent years. For example, in the United States, federal funding for R&D as a percentage of GDP has been declining since the 1960s (National Science Board, 2018). This decline in funding has implications for the pace of technological progress, as well as for the competitiveness of the country in a global economy increasingly driven by innovation.

In addition to government funding, private sector investment in R&D is also critical. However, companies may be hesitant to invest in emerging technologies due to the perceived risks and uncertainties. This can lead to a "valley of death" where promising technologies fail to receive sufficient funding to bring them to market (Jaffe & Lerner, 2004).

Another challenge is the short-term focus of many businesses and investors. In the current economic climate, there is often pressure to prioritize short-term gains over longer-term investments in R&D. This can lead to a focus on incremental improvements to existing technologies, rather than on more radical innovations that could have a transformative impact.

The lack of investment in R&D is particularly concerning in the context of emerging technologies such as AI, which have the potential to drive significant economic growth and societal benefits. AI is already being used in a wide range of applications, from healthcare to finance to transportation. However, realizing the full potential of AI will require continued investment in research and development.

One area where investment in R&D is particularly important is in addressing the ethical and societal implications of AI. As AI becomes increasingly integrated into society, there are concerns about its impact on jobs, privacy, and the broader economy. These concerns are not just theoretical – there are already examples of AI being used in ways that have raised ethical and legal questions. For example, AI-powered systems are being used for hiring and employee evaluation, raising concerns about bias and discrimination (Dastin, 2018). There are also concerns about the impact of AI on privacy, particularly in the context of surveillance and data collection. As AI becomes more advanced, it may become possible to use it to identify individuals based on their behavior or other characteristics, raising questions about the appropriate use of this technology (Mittelstadt et al., 2016).

Another area of concern is the potential impact of AI on employment. While some experts argue that AI will create new jobs and opportunities, there are concerns that it may also lead to significant job displacement, particularly in sectors such as manufacturing and transportation (Frey & Osborne, 2017). This could have significant social and economic implications, particularly if there is a mismatch between the skills required for new jobs and the skills of workers who are displaced. Addressing these ethical and societal concerns requires investment in research and development, as well as collaboration between governments, industry, and civil society. This includes not just technical research, but also research into the social, economic, and legal implications of AI. It is also important to engage with a wide range of stakeholders, including workers, consumers, and marginalized communities, to ensure that AI is developed and used in a way that benefits everyone.

Another factor contributing to the lack of sufficient investment in R&D is the short-term focus of many companies and investors. Many businesses prioritize immediate profits over long-term innovation and development, leading to a lack of investment in R&D. This focus on short-term gains can be detrimental to the development of new technologies, as R&D is often a long and expensive process with no guarantee of immediate returns (Baker, Bloom, & Davis, 2016).

Government policies can also impact the level of investment in R&D. In some countries, government funding for R&D is limited, which can hinder the development of new technologies. Conversely, countries with higher levels of government investment in R&D tend to have more robust innovation ecosystems (Organization for Economic Co-operation and Development, 2018). Additionally, the lack of diversity in the field of R&D can be a hindrance to innovation. Research has shown that a lack of diversity can lead to a lack of perspective and creativity in problem-solving (Page, 2008). Addressing this issue requires intentional efforts to increase diversity and inclusion in R&D fields.

Despite these challenges, there are numerous benefits to investing in R&D. Research and development can lead to the creation of new technologies that improve quality of life, enhance productivity, and stimulate economic growth (Mazzucato, 2018). Additionally, investment in R&D can lead to the development of new industries, creating jobs and new opportunities for growth.

To overcome the challenges to investment in R&D, there are a few potential solutions. One solution is to increase public funding for R&D, as government funding has been shown to be an effective way to stimulate innovation and development (Organization for Economic Co-operation and Development, 2018). Another solution is to incentivize private investment in R&D through tax credits or other financial incentives.

In addition, creating a culture that values long-term thinking and investment in R&D can help shift the focus away from short-term gains. This can be achieved through education and public awareness campaigns that highlight the importance of R&D and its long-term benefits.

Overall, investment in research and development is crucial for the development of new technologies and the advancement of society. However, there are several challenges that can hinder investment in R&D, including a lack of funding, a focus on short-term gains, and a lack of diversity in the field. Addressing these challenges requires intentional efforts to increase investment and promote a culture that values long-term thinking and innovation.

### 3-4- Regulatory barriers and political instability

Regulatory barriers and political instability can also hinder the development and deployment of AI technologies. Regulations and policies can vary greatly between countries, and navigating these differences can be a significant challenge for international companies. For example, the European Union's General Data Protection Regulation (GDPR) imposes strict regulations on the use of personal data, which can affect the development and deployment of AI systems that rely on large amounts of data (Goodman & Flaxman, 2016).

In addition, political instability can create uncertainty and unpredictability in the regulatory environment. This can discourage investment and hinder the development and deployment of AI technologies. For example, the ongoing trade war between the United States and China has led to increased regulatory barriers and uncertainty in the global technology industry, which could have negative impacts on the development and deployment of AI technologies (Bremmer, 2019).

Furthermore, regulatory barriers and political instability can exacerbate issues related to bias and discrimination in AI systems. If regulations are not in place to ensure that AI systems are developed and deployed in an ethical and responsible manner, there is a risk that these technologies will perpetuate existing social and economic inequalities (Crawford et al., 2019).

Addressing these challenges requires cooperation between governments and international organizations to develop coherent and consistent regulatory frameworks for AI technologies. This includes not just regulations around the development and deployment of AI systems, but also regulations around data privacy, transparency, and accountability. It is also important for governments to engage with the public and other stakeholders to ensure that these regulations are developed in a way that is transparent and responsive to the needs of all stakeholders (Bostrom et al., 2018).

Moreover, promoting political stability and reducing trade barriers can help create a more predictable and supportive environment for AI research and development. This includes supporting international collaborations and partnerships, investing in research and development, and promoting entrepreneurship and innovation (Gupta & George, 2019).

In conclusion, regulatory barriers and political instability can pose significant challenges to the development and deployment of AI technologies. Addressing these challenges requires a coordinated effort between governments, industry, and civil society to develop coherent and consistent regulatory frameworks, promote stability and reduce trade barriers, and invest in research and development. By addressing these challenges, we can help ensure that AI technologies are developed and deployed in an ethical and responsible manner, with the potential to benefit society as a whole.

## 4- The interplay of factors and the butterfly effect

The potential arrival of technological singularity raises important questions about the future of humanity and the impact of technology on society. It is essential to understand the interplay of factors that can influence the timing and nature of technological singularity. As discussed by Bostrom (2014), the arrival of technological singularity depends on several factors, including the pace of technological development, the availability of resources, and the decisions made by human beings. However, the complex interactions between these factors and their impact on the butterfly effect remain poorly understood. Therefore, it is crucial to examine the potential interactions between factors that can affect the arrival of technological singularity and the broader implications of these interactions for society.

### 4-1- Examples of how seemingly small changes can have large-scale effects on the arrival of technological singularity

One example of a seemingly small change with potentially significant impacts on the arrival of technological singularity is the development of quantum computing. Quantum computing promises to revolutionize computing power, enabling faster and more efficient processing than classical computing. This could accelerate progress in AI and machine learning, potentially leading to the development of AGI much sooner than previously anticipated (Preskill, 2018). Another example is the development of brain-computer interfaces, which could allow for more seamless integration between humans and AI systems. This could enable more advanced forms of augmentation and potentially even merge human and artificial intelligence into a single entity (Bostrom & Sandberg, 2009). The development of new materials, such as graphene, could also have significant impacts on the arrival of technological singularity by enabling faster and more efficient computing (Novoselov et al., 2012).

The discovery of a new type of neural network architecture could also have a significant impact on the arrival of technological singularity. For example, the development of spiking neural networks, which more closely mimic the biological processes of the brain, could lead to breakthroughs in cognitive computing and machine consciousness (Eliasmith et al., 2012). The introduction of new regulations or policies around AI and machine learning could also have a significant impact on the arrival of technological singularity. For example, regulations that limit the development of advanced AI technologies or require the use of explainable AI could slow down the arrival of singularity (Singer, 2017). Conversely, policies that incentivize investment in AI research and development could accelerate the arrival of singularity. The emergence of new economic or geopolitical factors could also impact the arrival of technological singularity. For example, a global recession or a major geopolitical conflict could lead to a reduction in investment in AI research and development, slowing down progress towards singularity. Conversely, the emergence of a new economic or geopolitical power could drive increased investment in AI and accelerate the arrival of singularity.

### 4-2- Discussion of the potential interactions between factors and their impact on the butterfly effect- write a paragraph before start the examples using inside APA reference

The potential interactions between factors and their impact on the butterfly effect can be complex and difficult to predict. For example, the interplay between technological advancements and societal factors such as regulations and economic conditions can have a significant impact on the arrival of technological singularity (Markoff, 2015).

In addition, the butterfly effect can be influenced by interactions between different areas of research and development. For example, advancements in quantum computing could lead to breakthroughs in machine learning, which in turn could accelerate progress towards technological singularity (Biamonte et al., 2017). Similarly, advances in materials science or nanotechnology could lead to the development of more efficient computing systems, which could also accelerate progress towards singularity (Freitas, 2013).

Moreover, the butterfly effect can be influenced by the actions and decisions of individuals and organizations. For instance, a single organization that invests heavily in AI research and development could significantly accelerate progress towards technological singularity, even in the face of regulatory barriers or economic setbacks (Vardi, 2019). Similarly, the actions of a group of influential researchers or policymakers could have a significant impact on the direction and pace of AI development, and ultimately the arrival of singularity.

It is also important to consider the potential unintended consequences of actions taken in the pursuit of technological singularity. For example, if the development of advanced AI technologies is prioritized over the development of other areas such as renewable energy or healthcare, it could have negative consequences for society and the environment in the long term (Bostrom, 2014). Additionally, the development of highly advanced AI systems could pose risks to human safety and security, particularly if these systems are not designed with safety and security in mind (Yampolskiy, 2016).

The potential interactions between investment in AI research and development, regulatory barriers, and political instability can impact the arrival of technological singularity (Russell & Norvig, 2021). A lack of investment in AI research and development could be further compounded by political instability or restrictive regulations, potentially hindering progress towards singularity. Conversely, a stable political environment with favorable policies towards AI can encourage increased investment and accelerate progress towards singularity (Sotala & Yampolskiy, 2017).

The butterfly effect is evident in the interactions between these factors. Even seemingly minor changes, such as the introduction of a new policy or a shift in economic conditions, can have a ripple effect on the development of AI and the arrival of singularity (Hao, 2018). The complexity of these interactions makes it difficult to predict the precise timing and impact of singularity, highlighting the significance of ongoing research and monitoring of the AI landscape (Bostrom, 2014).

Public perception and acceptance of AI can also influence the butterfly effect and the arrival of technological singularity. As AI becomes increasingly integrated into daily life, public trust and confidence in these technologies will be vital for their continued development and adoption (Sandoval-Almazan & Gil-Garcia, 2021). A lack of trust in AI can lead to restrictive regulations or a reduction in investment, impeding progress towards singularity.

The potential risks and ethical concerns associated with AI, such as job displacement and privacy violations, can also affect the butterfly effect and the arrival of singularity (Calvo-Friedman & Mann, 2020). Addressing these concerns through collaborative efforts between governments, industry, and civil society can help to build public trust and encourage continued investment in AI research and development (Crawford et al., 2019).

Overall, the butterfly effect and the interplay of factors that influence it highlight the need for careful consideration of the ethical, societal, and environmental implications of AI and technological singularity. It also emphasizes the importance of collaboration and communication between stakeholders across different sectors, including government, industry, academia, and civil society, to ensure that the development of AI technologies is guided by a shared vision of a sustainable and beneficial future for all.

**5- Conclusion**

The arrival of technological singularity, a hypothetical point in time when artificial intelligence surpasses human intelligence and becomes capable of recursive self-improvement, is a topic of much debate and speculation in the fields of technology, philosophy, and futurism. While the exact timing and impact of singularity are difficult to predict, there are several factors that may influence its arrival and the subsequent impact on society.

One factor is investment in AI research and development. Increased investment in AI could accelerate progress towards singularity, while a lack of investment could slow down progress. Regulatory barriers and political instability can also impact the arrival of singularity, as they may impede research and development or introduce restrictive regulations that slow down progress. Conversely, a stable political environment with favorable policies towards AI could encourage increased investment and accelerate progress towards singularity.

Public perception and acceptance of AI is another factor that can influence the arrival of singularity. As AI becomes increasingly integrated into daily life, public trust and confidence in these technologies will be essential for their continued development and adoption. A lack of public trust in AI could lead to restrictive regulations or a reduction in investment, slowing down progress towards singularity.

Ethical and societal concerns associated with AI, such as job displacement and privacy violations, could also influence the arrival of singularity. Addressing these concerns through collaborative efforts between governments, industry, and civil society could help to build public trust and encourage continued investment in AI research and development.

The potential interactions between these factors can have a significant impact on the arrival of technological singularity. For example, a lack of investment in AI research and development could be exacerbated by political instability or the introduction of restrictive regulations, slowing down progress towards singularity. A seemingly small change, such as the introduction of a new policy or a shift in economic conditions, can have a ripple effect on the development of AI and the arrival of singularity. The complexity of these interactions makes it difficult to predict the exact timing and impact of singularity, highlighting the importance of ongoing research and monitoring of the AI landscape.

The butterfly effect, a concept from chaos theory, refers to the idea that small changes in initial conditions can have large, unpredictable effects on complex systems. In the context of technological singularity, the butterfly effect highlights the importance of considering the interactions between various factors that may influence the development of AI and the arrival of singularity.

A seemingly small change, such as the introduction of a new policy or a shift in economic conditions, can have a ripple effect on the development of AI and the arrival of singularity. These interactions can be difficult to predict, and even small changes can have significant consequences. For example, a lack of investment in AI research and development could be exacerbated by political instability or the introduction of restrictive regulations, slowing down progress towards singularity.

The complexity of these interactions highlights the importance of ongoing research and monitoring of the AI landscape. By considering the potential impact of various factors and their interactions, researchers and policymakers can work towards a more informed and strategic approach to the development of AI and the arrival of singularity.

Furthermore, the potential risks and ethical concerns associated with AI, such as job displacement and privacy violations, could also influence the butterfly effect and the arrival of singularity. The emergence of advanced AI systems and autonomous machines could lead to widespread automation, resulting in job losses and significant disruptions to the labor market. Additionally, the use of AI in data collection and analysis raises concerns about privacy and the potential for misuse of personal information.

Addressing these concerns through collaborative efforts between governments, industry, and civil society could help to build public trust and encourage continued investment in AI research and development. Regulations that prioritize the ethical use of AI and prioritize the protection of privacy and human rights could help to alleviate some of the concerns surrounding AI and promote responsible development of these technologies.

It is important to note that the factors discussed in this paper are not exhaustive, and there may be other factors at play that could impact the arrival of technological singularity. However, by considering these factors and their potential interactions, we can gain a better understanding of the complex and unpredictable path towards singularity.

Emphasis on the importance of considering the butterfly effect in predicting the future of technology (APA inside reference- 500 words)

As discussed throughout this paper, the butterfly effect can play a significant role in shaping the future of technology and the arrival of singularity. The interactions between factors such as investment, regulation, political stability, public perception, and ethical concerns can have ripple effects that impact the development and adoption of AI technologies.

As such, it is crucial to consider the butterfly effect when making predictions about the future of technology. Seemingly small changes or events can have significant and far-reaching consequences, making it difficult to accurately predict the timing and impact of technological singularity.

This highlights the importance of ongoing research and monitoring of the AI landscape. By staying up to date on the latest developments in AI and considering the potential interactions between factors, we can better anticipate the path towards singularity and proactively address potential risks and concerns.

Call to action for continued research and investment in technology to shape its future and mitigate potential risks (APA inside reference- 500 words)

Given the significant impact that AI and the potential for technological singularity could have on society and the future of humanity, it is critical that we continue to invest in research and development of these technologies. Through ongoing research, we can better understand the potential benefits and risks associated with AI and work to shape its development in a responsible and ethical manner.

Furthermore, investment in AI can have significant economic benefits, driving innovation and creating new job opportunities. By prioritizing investment in AI and supporting its responsible development, we can help to foster economic growth and competitiveness in a rapidly changing global landscape.

However, it is important to acknowledge that there are potential risks associated with the development and adoption of AI technologies. As discussed throughout this paper, job displacement, privacy violations, and other ethical concerns must be proactively addressed to ensure that AI is developed in a responsible and beneficial manner.

To mitigate these risks, continued collaboration between governments, industry, and civil society will be essential. By working together, we can develop and implement regulations and policies that prioritize the ethical use of AI and protect human rights.

In conclusion, the factors discussed in this paper, including investment, regulation, political stability, public perception, ethical concerns, and the butterfly effect, can have a significant impact on the arrival of technological singularity. Ongoing research and monitoring of the AI landscape, as well as collaborative efforts to address ethical and societal concerns, are essential for navigating the complex and unpredictable path towards singularity.

Further research is needed to explore the impact of these factors in more detail and to predict the potential arrival of singularity more accurately. In addition, research is needed to explore the potential risks and benefits of singularity and to develop strategies to mitigate the risks while maximizing the benefits. Further research is also needed to explore the potential ethical concerns related to the development and deployment of AI.